\documentclass[a4paper,11pt]{article}
\usepackage{pos}

\title{The Deep Underground Neutrino Experiment (DUNE) program}

\author*{Inés Gil-Botella}
\onbehalf{on behalf of the DUNE collaboration}

\affiliation{CIEMAT, Centro de Investigaciones Energéticas Medioambientales y Tecnológicas,\\ 
Av. Complutense, 40 - E-28040 Madrid, Spain}

\emailAdd{ines.gil@ciemat.es}

\abstract{The Deep Underground Neutrino Experiment (DUNE) is a next-generation long-baseline neutrino oscillation experiment aimed at determining the neutrino mass hierarchy and the CP-violating phase. The DUNE physics program also includes the detection of astrophysical neutrinos and the search for signatures beyond the Standard Model, such as nucleon decays. DUNE consists of a near detector complex located at Fermilab and four 17 kton Liquid Argon Time Projection Chamber (LArTPC) far detector modules to be built 1.5 km underground at SURF, approximately 1300 km away. The detectors are exposed to a wideband neutrino beam generated by a 1.2 MW proton beam with a planned upgrade to $>$ 2 MW. Two 770 ton LArTPCs (ProtoDUNEs) have been operated at CERN for over 2 years as a testbed for DUNE far detectors and have been optimized to take new cosmic and test-beam data in 2024-2025. The DUNE and ProtoDUNE experiments and physics goals, as well as recent progress and results, are presented.}

\FullConference{42nd International Conference on High Energy Physics (ICHEP2024)\\
18-24 July 2024\\
Prague, Czech Republic\\}

\begin{document}
\maketitle



\section{The DUNE physics program}


The Deep Underground Neutrino Experiment (DUNE) \cite{DUNE:2020lwj} is a world-leading liquid argon TPC (LArTPC) neutrino oscillation experiment with the potential to deliver groundbreaking results as the unambiguous determination of the mass ordering, and the discovery of leptonic CP violation. The DUNE physics program also includes the detection of astrophysical neutrinos and the search for physics Beyond the Standard Model (BSM).


DUNE will detect neutrinos from the most powerful neutrino beam in the world ($>$ 2 MW) produced at Fermilab, and sent to the Sanford Underground Research Facility in Lead, South Dakota (SURF) along 1300 km of distance. A suit of four Far Detector (FD) modules (70 kton LAr TPCs) will be installed 1.5 km underground at SURF to observe the neutrino interactions. A Near Detector (ND) complex at 560 m from the source will measure the neutrino flux and provide an unprecedented control of the systematic uncertainties to enable precision studies of neutrino oscillations.

\begin{figure}[htb]
    \centering
    \includegraphics[width=0.49\textwidth]{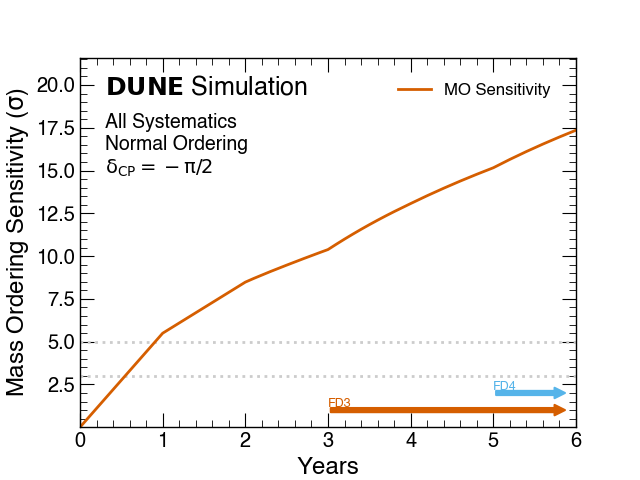}
    \includegraphics[width=0.49\textwidth]{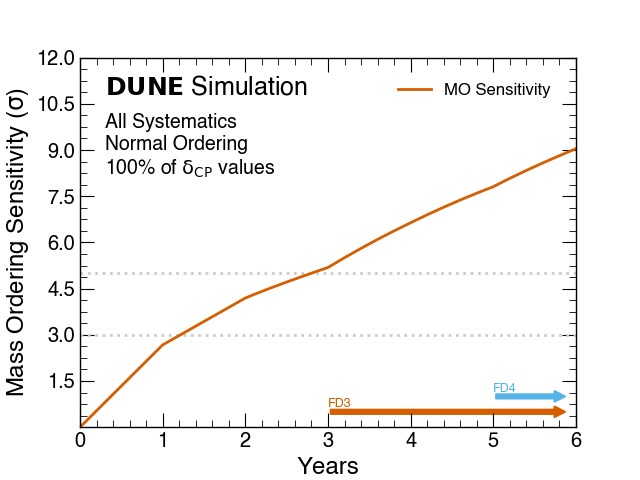}
    \caption{The significance for DUNE to establish the mass ordering for the best case (left panel) and worse case (right) as a function of running time.}
    \label{fig:mo}
\end{figure}
\begin{figure}[htb]
    \centering
    \includegraphics[width=0.49\textwidth]{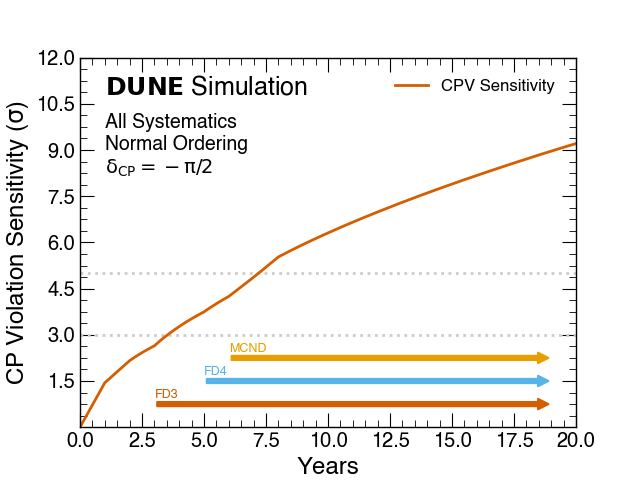} 
    \includegraphics[width=0.49\textwidth]{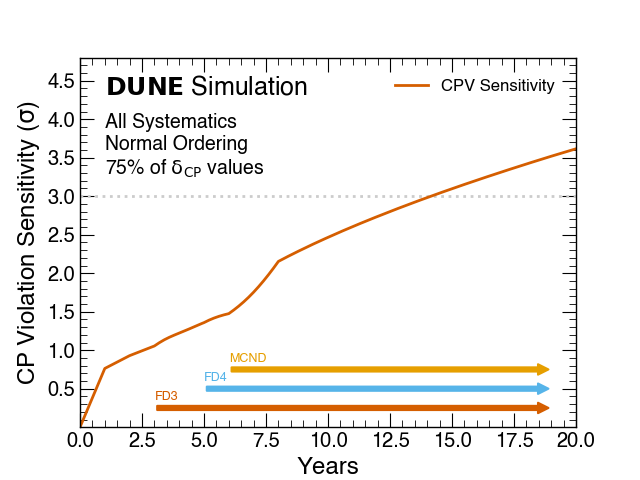}
    \caption{The significance for DUNE to establish CP violation for the best case (left panel) and 75\% (right) of $\delta_{CP}$ values as a function of running time.}
    \label{fig:cpv}
\end{figure}

The measurement of the neutrino and antineutrino energy spectra at the FD modules will enable the determination of the CP phase, neutrino mass ordering and $\theta_{23}$ mixing angle \cite{DUNE:2020jqi, DUNE:2021mtg}. Figures \ref{fig:mo} and \ref{fig:cpv} show the DUNE neutrino mass ordering and CP violation sensitivity as a function of the years for the currently assumed staging scenario. For the best case ($\delta_{CP}$ = -$\pi$/2), DUNE will reach a 5$\sigma$ mass ordering sensitivity in 1 year of data and 3$\sigma$ in 3.5 years. For the worst case, DUNE will need only 3 years to reach 5$\sigma$ sensitivity for the mass ordering and in long-term, DUNE can establish CP violation over 75\% of the $\delta_{CP}$ values at more than 3$\sigma$.

\begin{figure}[htb]
    \centering
    \includegraphics[width=0.33\textheight]{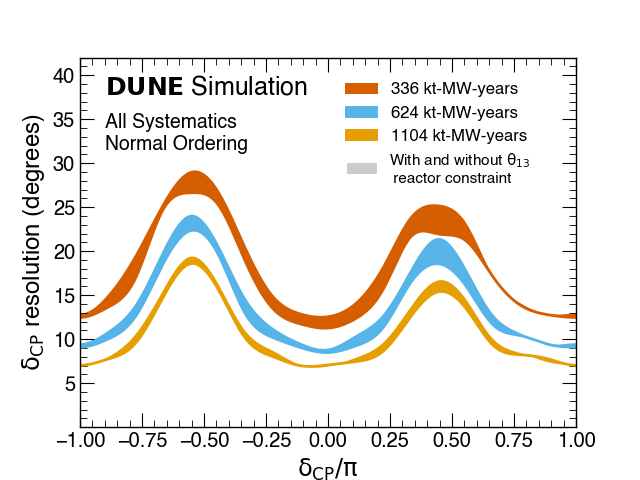}
    \includegraphics[width=0.3\textheight]{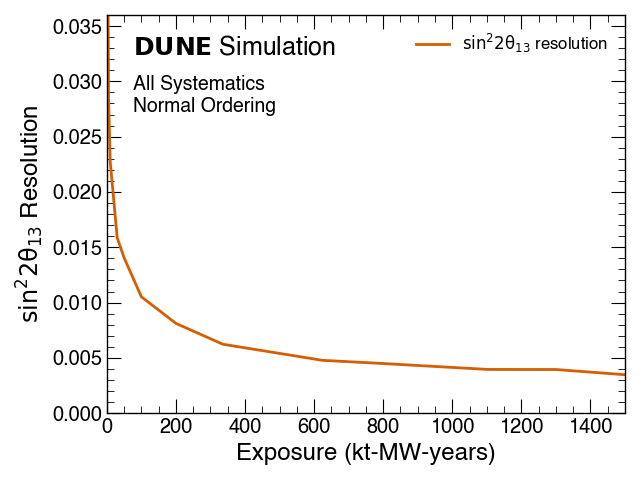}
    \caption{(Left panel) The DUNE resolution to $\delta_{CP}$ for as a function of the true $\delta_{CP}$ value for different integrated exposures. (Right panel) The DUNE resolution to sin$^22\theta_{13}$ as a function of exposure in kt-MW-yrs.}
    \label{fig:exposure}
\end{figure}

The ultimate goal of DUNE is to provide high precision measurements of the oscillation parameters in a single experiment. DUNE will reach between 6-16 degrees resolution to $\delta_{CP}$ (with external input for only solar parameters) depending on the true $\delta_{CP}$ value and world-leading precision in the $\theta_{13}$ measurement by a long-baseline experiment, to be compared with the $\theta_{13}$ measurement by future reactor neutrino experiments (Figure \ref{fig:exposure}). 

\begin{figure}[htb]
    \centering
    \includegraphics[width=0.58\textwidth]{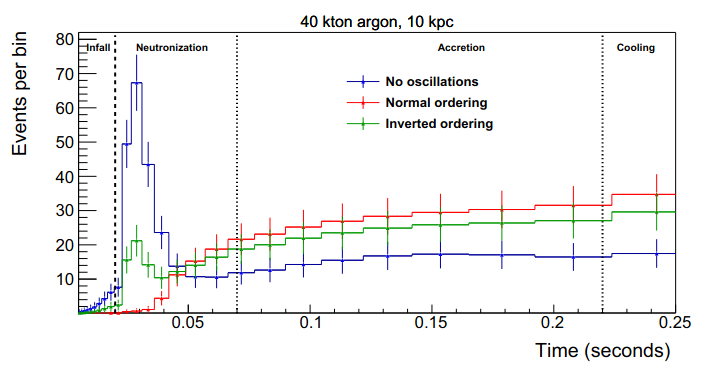}
    \includegraphics[width=0.4\textwidth]{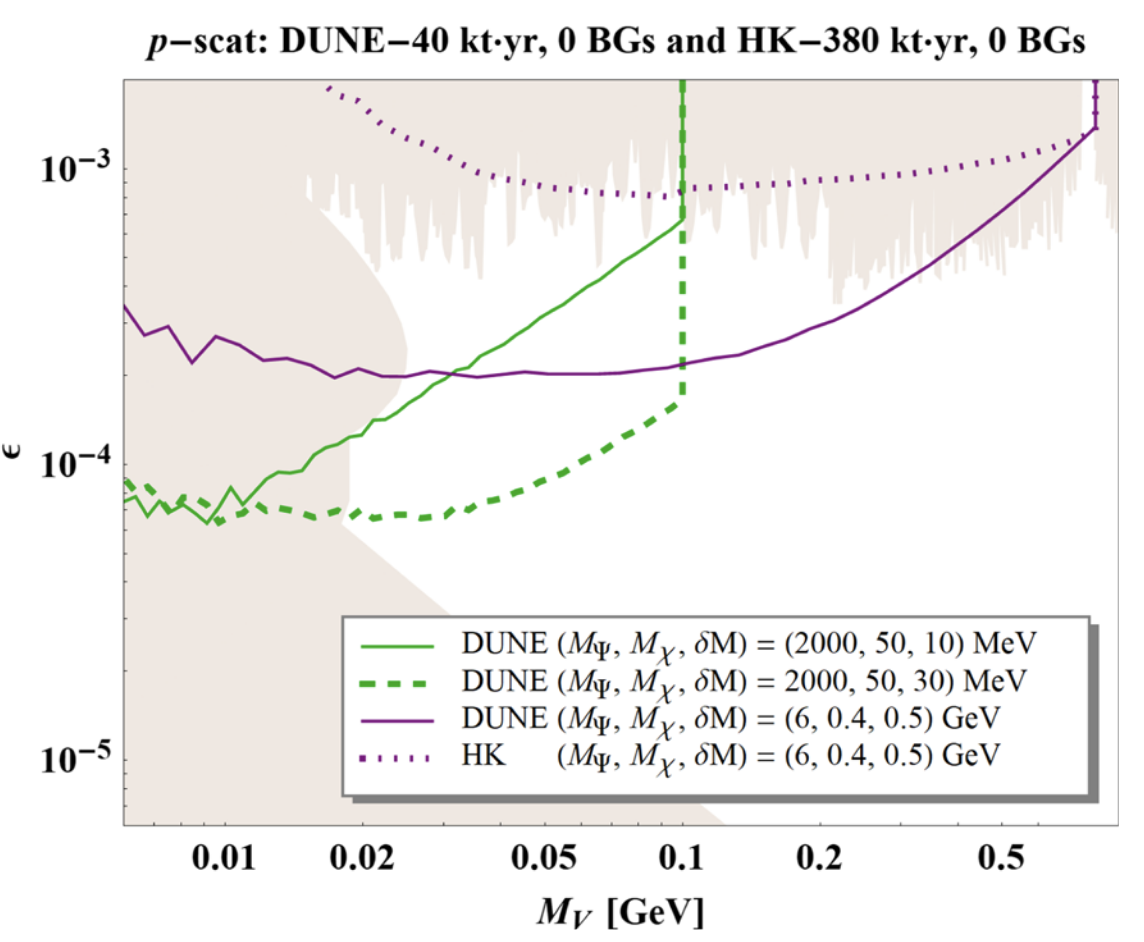}  
 \caption{Expected event rates as a function of time for 40 kton of argon during early stages of a supernova burst at 10 kpc (left). DUNE sensitivity to inelastic boosted dark matter search in the $p$-scattering channel \cite{DUNE:2020fgq} (right).} 
    \label{fig:solar-sn}
\end{figure}

Besides the long-baseline neutrino oscillation program, DUNE has a broad non-beam physics program. The DUNE underground location allows for astrophysical neutrino measurements, in particular neutrinos from core-collapse supernovae and the sun. DUNE has unique sensitivity to MeV electron neutrinos, thanks to the $\nu_e$CC absorption channel on $^{40}$Ar. DUNE expects to detect several thousand events from a galactic supernova burst. As shown in Figure \ref{fig:solar-sn}-left, DUNE can detect the supernova neutronization burst, which is an expected early emission of $\nu_e$, and measure the neutrino mass ordering by looking the suppression or not of this burst \cite{DUNE:2020zfm}. DUNE is also sensitive to all neutrino flavors through the elastic-scattering (ES) channel, which provides pointing capabilities, being able to achieve $\sim$5 degrees pointing resolution \cite{DUNE:2024ptd}.

DUNE has the potential to detect a huge amount of electron neutrinos from the sun ($\sim$10k/year CC; 2k ES/year) together with radioactive background events expanding up to $\sim$14 MeV (mainly neutron-capture events from the cavern surrounding the detector at SURF). The $^8B$ solar neutrino flux is visible above 9-10 MeV and the hep flux can be measured for the first time. The measurement of the $^8B$ neutrinos will improve current solar measurements of $\Delta m^2_{21}$ because of the day-night asymmetry induced by Earth matter effects. 

DUNE provides enormous opportunities to probe phenomena beyond the Standard Model (BSM) \cite{DUNE:2020fgq}. 
By measuring neutrino and antineutrino spectra, DUNE can look for 
neutrino oscillation scenarios beyond the standard three-flavor picture, including sterile neutrinos, and PMNS non-unitarity.
The high statistics in neutrino and antineutrino measurements allow for CPT violation searches. DUNE has unique sensitivity to non-standard interactions due to long-baseline matter effects. Other BSM searches in ND and FD are possible such as the search for inelastic dark matter scattering at the FD (Figure \ref{fig:solar-sn}-right), low-mass dark matter at the ND, nucleon decay, neutron-antineutron oscillations, heavy-neutral leptons, neutrino tridents, among others.

\section{The DUNE detectors}

\begin{figure}[htb]
    \centering
    \includegraphics[width=0.28\textwidth]{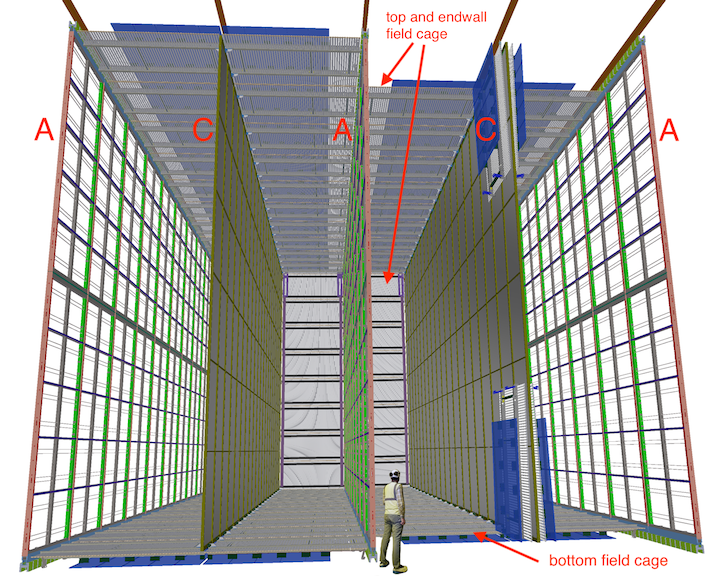}
    \includegraphics[width=0.33\textwidth]{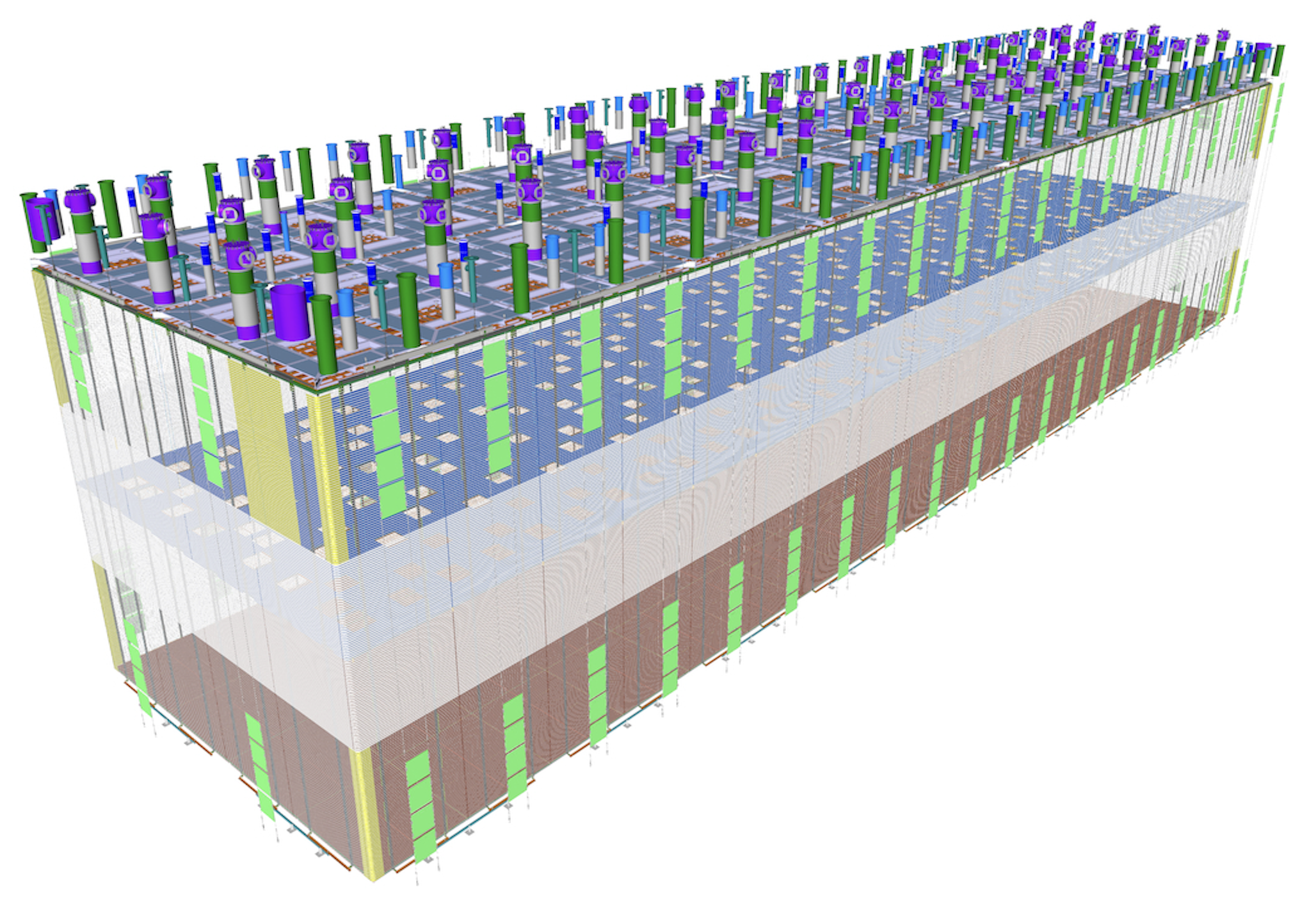}
    \includegraphics[width=0.37\textwidth]{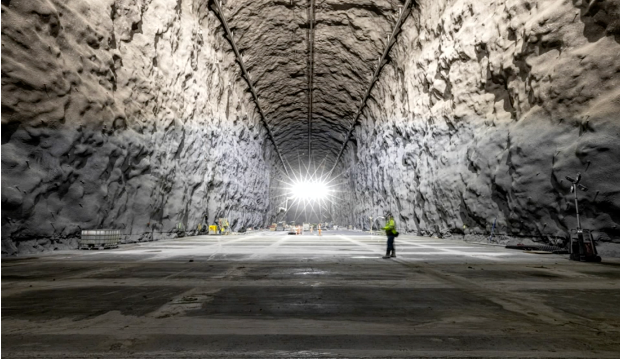}
    \caption{The horizontal drift FD-HD module (left), vertical drift FD-VD module (center), and one of the SURF caverns that will host the DUNE far detectors (right).}
    \label{fig:fd}
\end{figure}

To accomplish this physics program, DUNE is currently building two LArTPC FD modules. One module (FD-HD) has an horizontal drift (3.6 m drift) and uses wire readout planes with 4 drift regions (Figure \ref{fig:fd}-left) \cite{DUNE:2020txw}. In the other FD (FD-VD) \cite{DUNE:2023nqi}, the ionizing electrons drift vertically in two 6 m drift regions (Figure \ref{fig:fd}-center). This second technology is simpler to install and is the baseline for the third and forth FD modules. New caverns to allocate these huge detectors have been excavated at SURF (Figure \ref{fig:fd}-right).

The DUNE ND complex \cite{DUNE:2021tad} will enable the prediction of the FD reconstructed spectra. The ND will use a movable detector system composed of a LAr TPC (ND-LAr) and a muon spectrometer (TMS). They will collect high statistics of the neutrino beam. The off-axis data of this system will allow to constrain the energy dependence of the neutrino cross sections. The use of the same target and technology will inform predictions of the neutrino reconstructed energy in the FD. The final component of the ND complex will be an on-axis magnetized detector (SAND) to monitor the beam stability and perform measurements of neutrino interactions.


DUNE has developed a two-phase strategy toward the implementation of the leading-edge, large-scale science projects.
The construction of the initial experiment configuration (DUNE Phase I) is well underway. It includes the two FD modules and the ND components already described, with a 1.2 MW proton beam. DUNE Phase-II \cite{DUNE:2024wvj} consists of a third and fourth FD module, an upgraded ND complex, and an enhanced 2.1 MW beam. 

\section{The DUNE prototypes}

\begin{figure}[htb]
    \centering
    \includegraphics[height=0.18\textheight]{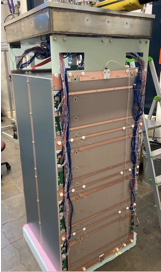}
    \includegraphics[height=0.18\textheight]{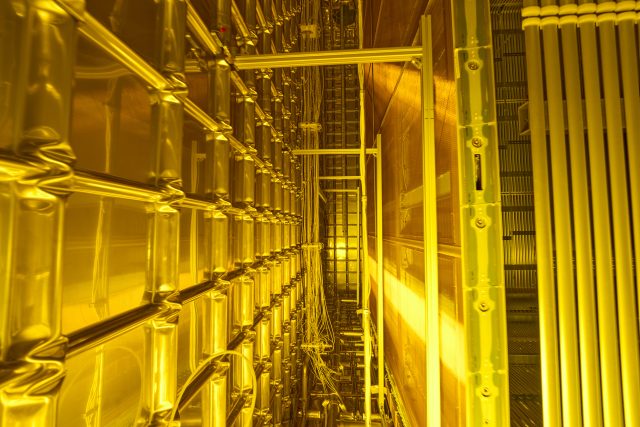}
    \includegraphics[height=0.18\textheight]{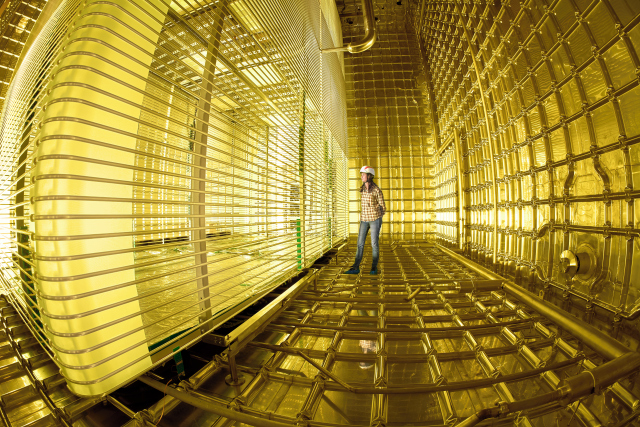}
    \caption{One of the four modules of the ND-LAr 2x2 demonstrator (left panel). Inner view of ProtoDUNE-HD (center) and ProtoDUNE-VD (right) at the CERN Neutrino Platform.}
    \label{fig:protodune}
\end{figure}

The DUNE collaboration is developing a very active prototyping program for both ND and FDs at Fermilab and CERN in order to test at large scale the main components and demonstrate the detector performance. In particular, one key aspect of the ND-LAr operation is the ability to cope with a large number of neutrino interactions in each spill. The expected high rate at the near site motivates the use of a pixelated readout and optical modularity in the ND. In order to test this new technology, four LAr TPC modules (Figure \ref{fig:protodune}-left) have been built and operated in LAr at the University of Bern and then moved to Fermilab and installed in the former location of the MINOS ND. The system also includes upstream and downstream MINERvA scintillator and calorimeter planes. The ND-LAr 2x2 demonstrator has detected the first neutrino interactions and will continue collecting data with the goal of proving the 3D readout in a neutrino beam with similar event rate to DUNE. 

In order to demonstrate the technical feasibility of the FDs and show the expected performance, two full-scale LAr prototypes (770 ton each) were built and operated at CERN between 2018 and 2020 \cite{DUNE:2021hwx}. After a successful operation with cosmic muons and charged particles, 
ProtoDUNEs are now entering a new phase. Two new prototypes were installed at the CERN Neutrino Platform, incorporating the final technical solutions for FD-HD and FD-VD (Figure \ref{fig:protodune}-center, right). ProtoDUNE-HD was filled with LAr and started taking cosmic rays and test-beam data at CERN in May 2024. This new run will increase beam data statistics for cross section measurements, particle identification, calibration, and reconstruction developments.
After the operation of ProtoDUNE-HD, LAr will be transferred to ProtoDUNE-VD in fall 2024 for running in early 2025.

The VD technology is the baseline design for the DUNE FD Phase II modules. Some improvements to the light collection for FD3 are already being explored. But the DUNE phased construction allows for more ambitious detector improvements that allow to expand the DUNE physics scope, in particular, the low-energy physics program. The collaboration is pursuing an active R\&D program for the so-called Module of Opportunity FD4 \cite{DUNE:2024wvj}, including pixel readout in the FD, integrated charge-light readout, as the SoLAr proposal \cite{SoLAr:2024fwt}, low background modules or non-LAr technologies.



\end{document}